\documentclass[reqno, 11pt,a4paper]{amsart}

\pdfoutput=1
\usepackage{calc,graphicx,amsfonts,amsthm,amscd,epsfig,psfrag,amsmath,amssymb,enumerate,dsfont}
\usepackage{subfig} 
\usepackage{pdfsync}
\usepackage{stmaryrd} 
\usepackage{cite}

\usepackage[colorlinks=true, pdfstartview=FitV, linkcolor=blue, citecolor=blue, urlcolor=blue,pagebackref=false]{hyperref}

\DeclareMathSymbol{\leqslant}{\mathalpha}{AMSa}{"36} 
\DeclareMathSymbol{\geqslant}{\mathalpha}{AMSa}{"3E} 
\DeclareMathSymbol{\eset}{\mathalpha}{AMSb}{"3F}     
\renewcommand{\leq}{\;\leqslant\;}                   
\renewcommand{\geq}{\;\geqslant\;}                   

\renewcommand{\r}{\rho}

\newcommand{\cM}{\ensuremath{\mathcal M}}

\newcommand{\cP}{\ensuremath{\mathcal P}}

\newcommand{\cZ}{\ensuremath{\mathcal Z}}

\newcommand{\bbE}{{\ensuremath{\mathbb E}} }

\newcommand{\bbR}{{\ensuremath{\mathbb R}} }

\let\m=\mu            
\let\r=\rho

\newcommand{\feta}{\boldsymbol{\eta}}

\newcommand{\rhotrans}{{\rho_{\rm trans}}}
\newcommand{\rhodyn}{{\rho_{\rm dyn}}}

\graphicspath{{./Figs/}}

\begin{document}
\author[P. Chleboun]{Paul Chleboun$^1$}

\author[S. Grosskinsky]{Stefan Grosskinsky$^{1,2}$}

\thanks{\noindent $^1$ Mathematics Institute, University of Warwick, Coventry CV4 7AL, UK.\\
\indent $^2$ Centre for Complexity Science, University of Warwick, Coventry CV4 7AL, UK.\\
\indent e--mail: paul@chleboun.co.uk, S.W.Grosskinsky@warwick.ac.uk}

\title{A dynamical transition and metastability in a size-dependent zero-range process}

\begin{abstract}
We study a zero-range process with system-size dependent jump rates, which is known to exhibit a discontinuous condensation transition. 
Metastable homogeneous phases and condensed phases coexist in extended phase regions around the transition, which have been fully characterized in the context of the equivalence and non-equivalence of ensembles. 
In this communication we report rigorous results on the large deviation properties and the free energy landscape which determine the metastable dynamics of the system. 
Within the condensed phase region we identify a new dynamic transition line which separates two distinct mechanism of motion of the condensate, and provide a complete discussion of all relevant timescales. 
Our results are directly related to recent interest in metastable dynamics of condensing particle systems.
Our approach applies to more general condensing particle systems, which exhibit the dynamical transition as a finite size effect.

\end{abstract}

\maketitle

\section{Introduction.}

The understanding of metastable dynamics associated to phase transitions in complex many-body systems is a classical problem in statistical mechanics. 
It is rather well understood on a heuristic level, characterizing metastable states as local minima of the free energy landscape and the transitions between them occurring along a path of least action, corresponding to the classical Arrhenius law of reaction kinetics \cite{hugoreview}. 
Since the classical work by Freidlin and Wentzell on random perturbations of dynamical systems \cite{freidlin}, there have been various rigorous approaches in the context of stochastic particle and spin systems summarized in \cite[Chapter 4]{Oliveira2005} and \cite{boviermeta}.
A mathematically rigorous treatment of metastability remains an intriguing question and is currently a very active field in applied probability and statistical mechanics \cite{hugo14,fernandez,cassandro,beltran13,landimasym}. 
Most recently, potential theoretic methods \cite{bovierlow} have been combined with a martingale approach to establish a general theory of metastability for continuous-time Markov chains \cite{beltran13,beltranetal10,beltranetal12b}.
The dynamics of condensation in driven diffusive systems has recently become an area of major research interest in this context.
There have also been recent results on the inclusion process \cite{jiarui,GRV} and systems exhibiting explosive condensation \cite{waclaw12,waclaw13}, however the zero-range process remains one of the most studied systems.


Zero-range processes (ZRPs) are stochastic lattice gases with conservative dynamics introduced in \cite{spitzer70}, and the condensation transition in a particular class of these models was established in \cite{drouffe98,evans00,stefan,godreche03}. 
Many variants of this class have been studied in recent years including a non-Markovian version with slinky condensate motion \cite{hirschberg12,hirschberg09}, see also \cite{godrecheetal12,CGreview} for a recent reviews of the literature. 
If the particle density $\rho$ exceeds a critical value $\rho_c$ the system phase separates into a homogeneous fluid phase at density $\rho_c$ and a condensate, which concentrates on a single lattice site and contains all the excess mass. 
The dynamics and associated timescales of this transition have been described heuristically in \cite{godrecheetal05}. 
For large but finite systems, due to ergodicity, the location of the condensate changes on a slow timescale and converges to a random walk on the lattice in the limit of diverging density \cite{beltranetal12}. 
Recent extensions of these rigorous results include a non-equilibrium version of the dynamics \cite{landimasym,beltranetal12b}, and a thermodynamic scaling limit with a fixed supercritical density $\rho >\rho_c$ \cite{AGL2}. 

While the motion of the condensate is the only metastable phenomenon in the above results, a slight generalization studied in \cite{Chleboun10,AGL,evansetal05b} exhibits metastable fluid states at supercritical densities, which are a finite-size effect and do not persist in the thermodynamic limit. 
The model we study here was first introduced in \cite{stefangunter} motivated by experiments in granular media, and is a zero-range process where the jump rates scale with the system size. 
This leads to an effective long-range interaction, and it is well known that these can give rise to metastable states that are persistent in the thermodynamic limit \cite{hugoreview}. 
The condensation transition in this ZRP is discontinuous with metastable fluid and condensed states above and below the transition density, respectively, and the model has a rich phase diagram.

As a first main contribution of this work we identify a new dynamic transition within the condensed phase region which separates two distinct mechanism of motion of the condensate. 
Secondly, we provide a complete discussion of all relevant timescales using a comprehensive approach in the context of large deviation theory, which proves to be a powerful tool for the characterization and study of phase transitions for nonequilibrium systems \cite{hugoreview,hugo14}. 
All results we report here are based on rigorous work which is presented in more detail in \cite{paulinprep}, and are applicable in a more genreal context. 
To our knowledge, this constitutes the first example of a condensing particle system that exhibits extended regions in phase space with coexisting metastable states, and is an important step to extend recent rigorous results on the condensate dynamics in such systems.

%
%
%

\section{Definitions and notation.}

We consider a zero-range process on a one dimensional lattice of $L$ sites with periodic boundary conditions. Particle configurations are denoted by vectors of the form $\feta = (\eta_x)_{x=1}^L$ where $\eta_x$ is the number of particles on site $x$, which can take any value in $\{0,1,2,\ldots\}$.
In the ZRP particles jump off a site $x$ with a rate that depends only on the number of particles on the departure site, and then move to another site $y$ according to a random walk probability $p(x,y)$, which we take to be of finite range and translation invariant.

The rate at which a particle exits a site is denoted by $g_L(\eta_x)$, where the system size dependence of the jump rates is indicated by the subscript.
We consider simple jump rates introduced in \cite{stefangunter} of the form
\begin{align}
  g_L (n) :=
  \begin{cases}
    c & \textrm{if } n \leq a L ,\\
    1 & \textrm{if } n > a L,
  \end{cases}
  \quad \textrm{for }\ n \geq 1\quad  \textrm{and }\ g_L(0) = 0\,, 
\end{align}
for some $c >1$ and $a > 0 $.

It is well known  that ZRPs exhibit stationary distributions which factorize over lattice sites, see for example \cite{CGreview,stefangunter,evansetal05}.
It is convenient to introduce a prior distribution (or reference measure) which is stationary, that will be used to characterize the canonical and grand-canonical distributions after proper renormalization. 
The prior distribution is also size-dependent and given by
\begin{align*}
  \overline P_L(\feta) := \frac{1}{\overline Z_L}\prod_{x=1}^L w_L(\eta_x)e^{-\eta_x}\quad\textrm{with}\quad w_L(n) = \prod_{i=1}^n\frac{1}{g_L(i)} =
  \begin{cases}
    c^{-n} & \textrm{for } n{\leq} aL\\
    c^{-\lfloor aL \rfloor} & \textrm{for } n {>} aL
  \end{cases},
\end{align*}
where the empty product is taken to be unity and the normalisation factor is $\overline Z_L= \big(\sum_{n\geq 0} w_L(n)e^{-n} \big)^L$. 
The above weights $w_L$ are stationary for the ZRP, and the additional factor $e^{-\eta_x}$ is a convenient choice so that they can be normalised, which allows the interpretation of free energies as large deviation rate functions (cf.~\cite{hugoreview}). 
%

Since the dynamics are irreducible and conserve the total particle number, 
on a fixed lattice starting from any initial condition with a fixed number $N$ of particles the systems is ergodic. 
In the long time limit the distribution will converge to the corresponding canonical distribution $P_{L,N} :=\overline P_L (.|\sum_x \eta_x= N)$, which is given by a conditional version of the reference measure,
\begin{align*}
  P_{L,N}(\feta) = \begin{cases}
\frac{1}{Z_{L,N}}\overline P_L(\feta) & \textrm{if $\ \sum\limits_{x=1}^L\eta_x = N$}\\
0 & \textrm{otherwise}
\end{cases}
\ \textrm{ where } \  Z_{L,N} = \overline P_L\left(\sum_{x=1}^L\eta_x = N\right).
\end{align*}



\begin{figure}[t]
  \centering
  \mbox{
  \includegraphics[width=0.52\textwidth]{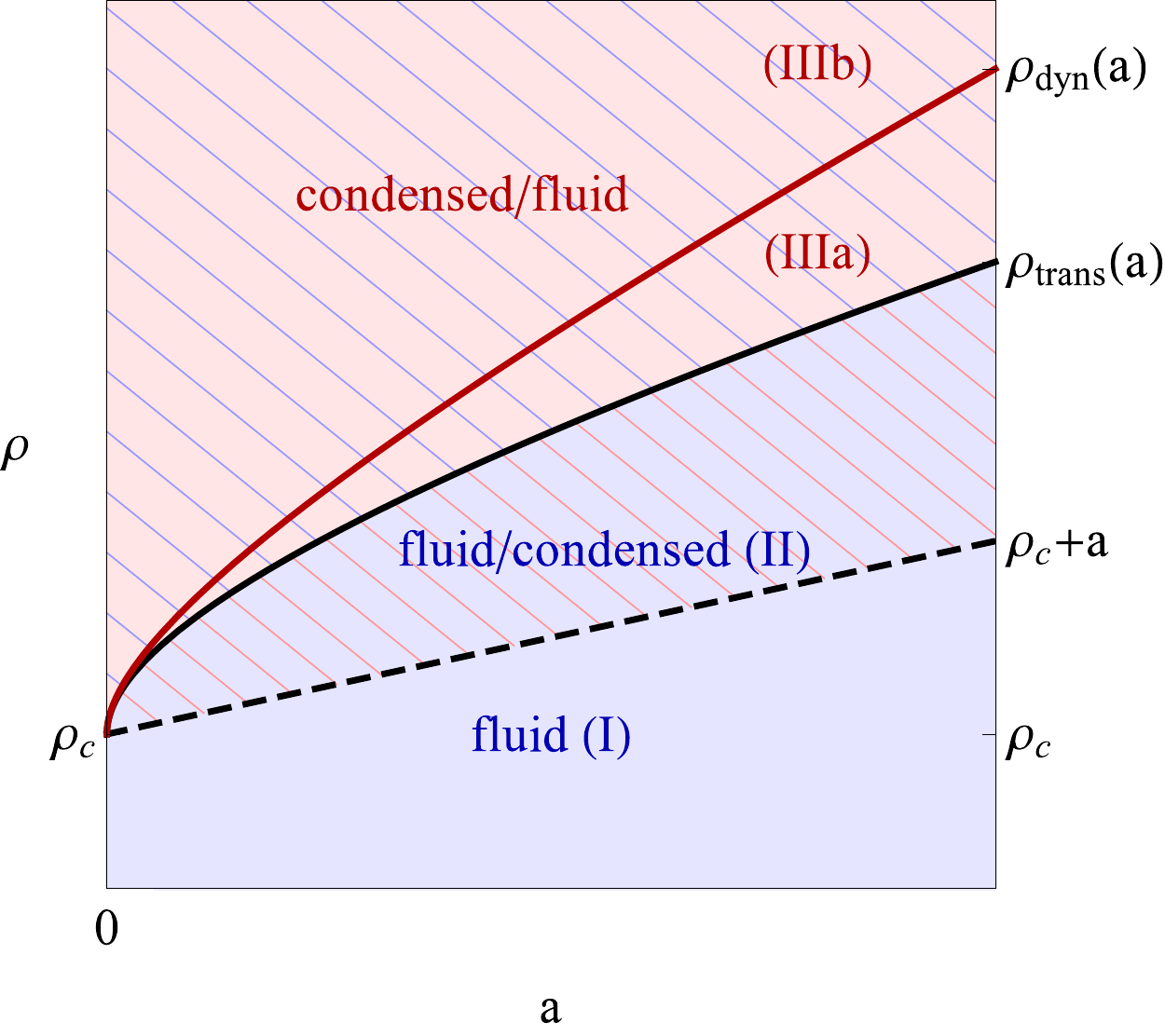}
  \quad
\includegraphics[width=0.4\textwidth]{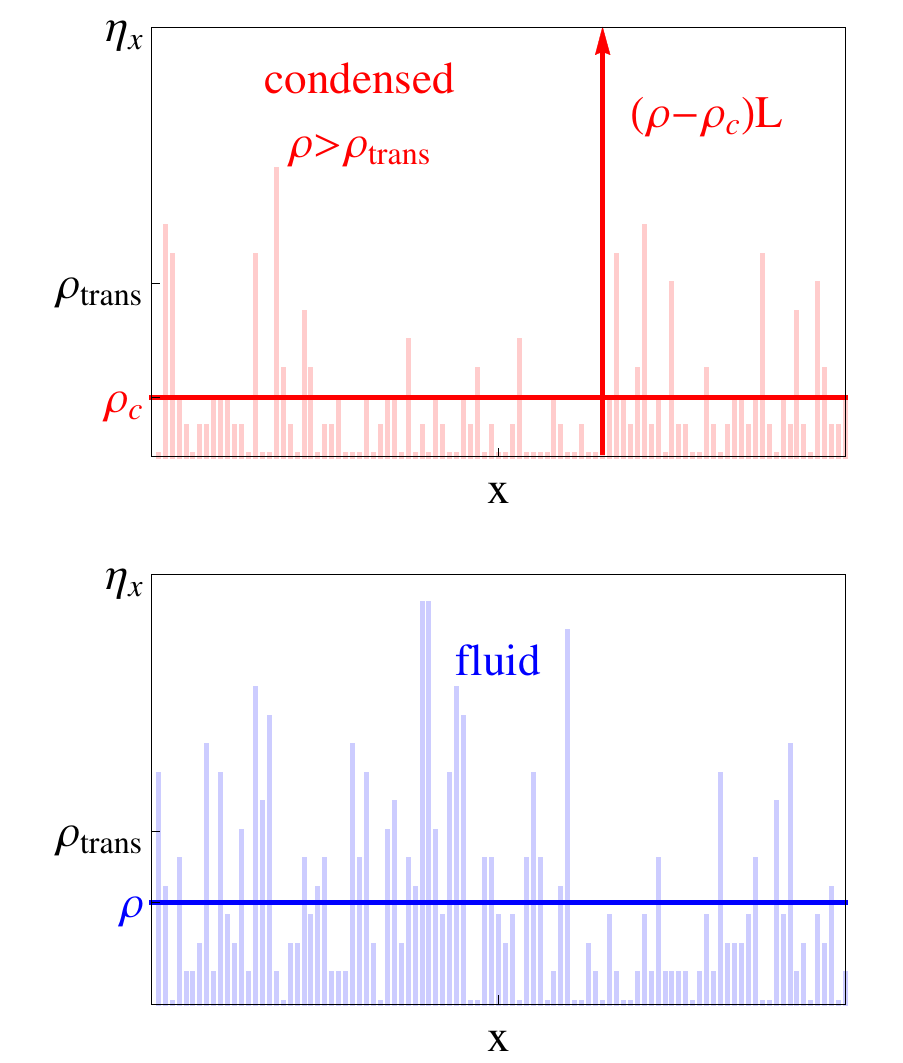}}
  \caption{\label{fig:configs} Phase diagram with the following phase regions: (I) For $\rho < \rho_c+a$ there is a unique fluid state and particles are distributed homogeneously. (II) For $\rho_c +a<\rho < \rhotrans$ an additional metastable condensed state exists. (III) For $\rho>\rhotrans$ the condensed state becomes stable, and the fluid state remains metastable for all densities. The new transition density $\rhodyn$ (\ref{eq:rhodyn}) characterizes a change in the mechanism for condensate motion, which is explained in Fig. \ref{fig:CartoonMotion}. Typical stationary configurations for fluid and condensed states are shown on the right.
}
\end{figure}

%



\section{Results}

The large scale behavior and condensation transition can be characterized as usual by the canonical free energy, which is defined as
\begin{align}\label{fdef}
  f(\rho) := -\lim_{\substack{L\to \infty\\ N/L \to \rho}}\frac{1}{L} \log Z_{L,N}\, .
\end{align}
Note that this is the large deviation rate function for the total number of particles under the reference measure $\overline P_L$  (cf. also \cite{hugoreview}) and is also the relative entropy of the canonical measures with respect to the reference distribution.
Explicit computation can be done using the grand-canonical and a restricted grand-canonical ensemble which we outline in the appendix.
In the following we simply report the main results. 
There exists a transition density $\rhotrans$ characterized in (\ref{eq:rhotra}), below which the system is typically in a fluid state with all particles distributed homogeneously. For $\rho >\rhotrans$, the system is in a condensed state, and phase separates into a single condensate site containing of order $(\rho -\rho_c )L$ particles and a fluid background at density $\rho_c$. As usual, this is characterized by the free energy decomposing into a contribution $f_{\rm fluid}$ from the fluid and $f_{\rm cond}$ from the condensed phase. It is given by
\begin{align}
f(\rho) =
	\begin{cases}
	f_{\rm fluid}(\rho) & \text{if } \rho \leq \rhotrans\quad\mbox{(fluid state)}\\
	f_{\rm fluid}(\rho_c) + f_{\rm cond} (\rho-\rho_c) & \text{if }  \rho > \rhotrans\quad\mbox{(condensed state)} .
	\end{cases}
\label{eq:free}
\end{align}
As derived in (\ref{eq:fldef}) in the appendix,
\begin{equation}
f_{\rm fluid} (\rho )=\rho\log\rho -(1+\rho )\log (1+\rho )+\rho\log (ce)-\log \Big( 1-\frac{1}{ce}\Big)\ ,
\label{eq:fflu}
\end{equation}
which is the relative entropy of a geometric distribution with density $\rho$ with respect to the reference measure. This geometric distribution can be interpreted as the fluid phase as is discussed in the appendix. 
The critical background density in the condensed phase is given in (\ref{eq:rhoc}) as $\rho_c =1/(c-1)$. 
The condensate contribution
\begin{align}
f_{\rm cond} (m)=\lim_{L\to\infty\atop M/L\to m} \frac{1}{L}\log\overline P_L (\eta_1 =M)=
\begin{cases}
  m + m\log c & \textrm{if } m < a,\\
  m + a\log c & \textrm{if } m \geq a
\end{cases}
\label{eq:fcond}
\end{align}
is determined by the reference probability of a single site containing $mL+ o(L)$ particles, since the condensate has no associated entropy. Note that even though fluid and condensate coexist in the condensed state, there is no free energy contribution from the interface since the stationary distributions factorize, and the combinatorial factor of $L$ possible positions for the condensate location only contributes on a sub-exponential scale. 
This lack of surface tension also implies that the condensed phase consists of a single site, in contrast to other systems with non-product stationary distributions \cite{evansetal06,waclawetal09}. 

The bevaviour of the free energy is dominated by typical stationary configurations, which are illustrated in Fig. \ref{fig:configs} along with the phase diagram of the model. 
Since the background density $\rho_c <\rhotrans$ is strictly smaller than the transition density the phase transition is a discontinuous transition, as opposed to condensation in ZRPs without size-dependent rates which was already reported in \cite{stefangunter}.
\\

\begin{figure}[t]
  \centering
  \mbox{
  \includegraphics[width=0.48\textwidth]{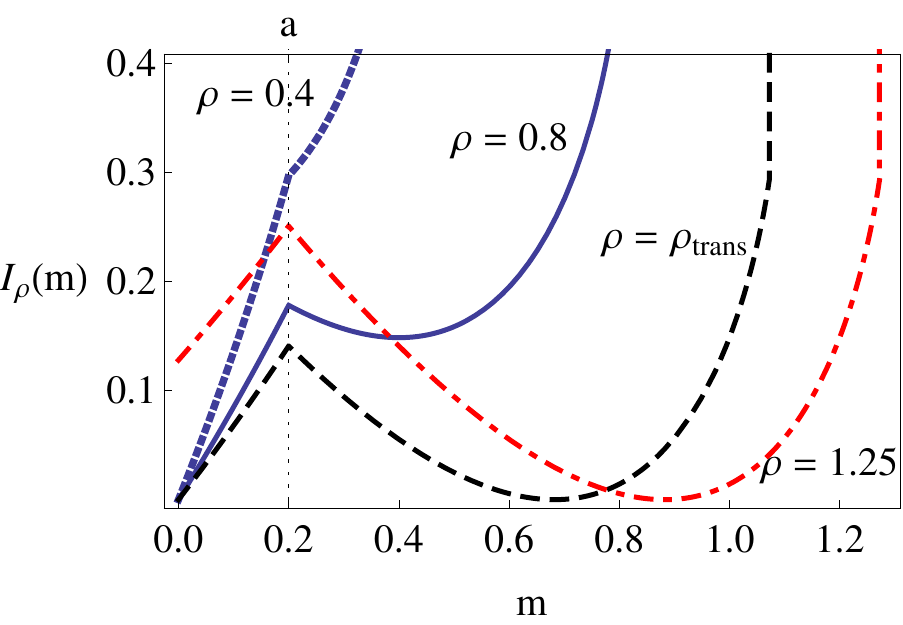}
  \quad
\includegraphics[width=0.48\textwidth]{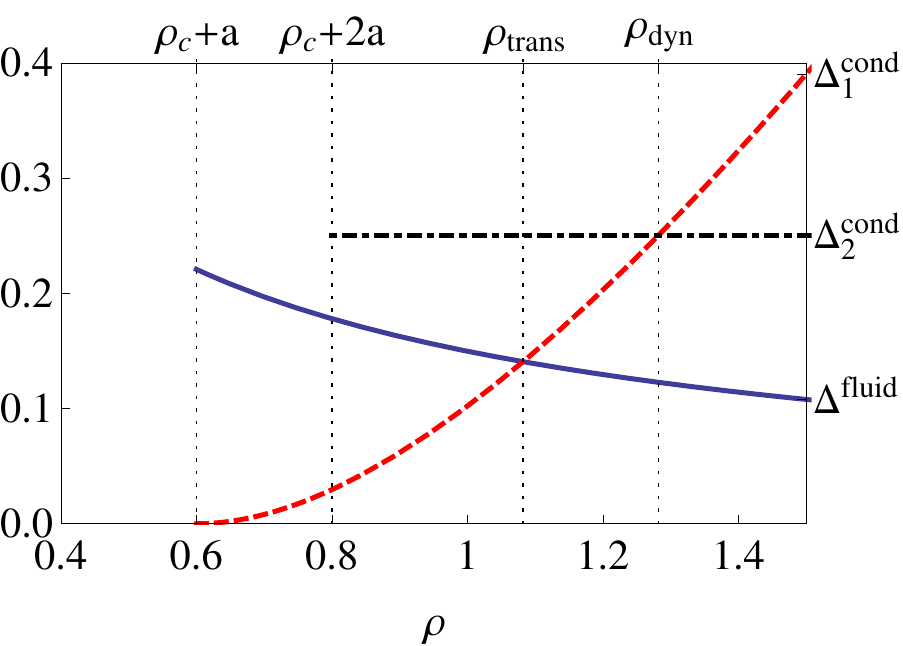}
  }
  \caption{\label{fig:ents} Left: The rate function $I_\rho (m)$ (\ref{eq:irhom}) for $a=0.2$, $c=3.5$ and various values of $\rho$, showing a single well for $\rho <\rho_c +a$ and a double well for higher densities. The change of global minimum happens at $\rho =\rhotrans$. 
  Right: The exponential rates of the lifetimes and condensate motion time, given by \eqref{eq:Deltas}, for $a=0.2$, $c=3.5$,
  characterizing the transition densities $\rho_{\rm tras}$ \eqref{eq:rhotra} and $\rho_{\rm dyn}$ \eqref{eq:rhodyn}.  
}
\end{figure}

\noindent\textbf{Metastable states.}\\
The phase diagram in \ref{fig:configs} also contains information about metastable states. They can be identified as local minima of the large deviation rate function  $I_{\rho}(m)$ for the maximum occupation number $\cM_L (\feta )=\max_{x} \eta_x$, as is shown in Fig. \ref{fig:ents}. 
This characterizes the exponential rate of decay of the canonical probability to observe a maximum of size $mL + o(L)$, i.e. 
\begin{align*}
  P_{L,N}\left( \cM_L  = M \right) \approx e^{- I_{\rho}(m)L}\quad\mbox{as }L\to\infty ,\ M/L\to m\ .
\end{align*}
In order to calculate $I_\rho(m)$, we first find joint large deviations of the maximum and the density under the prior distribution described by a rate function $I(\rho,m)$. 
Precisely, we can show that the following limit exists for $\rho > 0$ and $m \in [0,\rho]$
\begin{align*}
I(\rho ,m):=-\lim_{L\to \infty}\frac{1}{L} \log \overline P_L\left( \sum_{x=1}^L \eta_x = N,\ \cM_L = M\right) \quad\textrm{where} \quad N/L \to \rho, \ M/L \to m\ .
\end{align*}
The limit is independent of details of the sequences $N/L$ and $M/L$ so long as $m\in (0,\rho]$, if $M/L \to 0$ we require that $M$ is not too small.

We find that for each $\rho > 0$ and $m \in [0,\rho]$ the joint rate function for the density and maximum satisfies
 \begin{align}
 \label{eq:SizeDepIjoint}
   I(\rho,m) = 
  \begin{cases}
    f_{\rm cond} (m)+  f_{\rm fluid} (\rho-m) & \text{if } m < a \text{ or } m > (\rho - \rho_c),\\
    f_{\rm cond} (m) + \inf\limits_{m_2 \in[0,m]}I(\rho-m,m_2) & \text{otherwise,}  \\
  \end{cases}  
 \end{align}
and the iteration in the second case closes after finitely many steps for each $\rho< \infty$ and $m\in [0,\r]$.
The first term $f_{\rm cond} (m)$ is the contribution of the maximum to the rate function and the second term is the contribution due to the bulk of the system. 
The infimum in the second line of \eqref{eq:SizeDepIjoint} arise since a large deviation outside the range $m < a$ or $m> (\rho-\rho_c)$, which is always atypical and never locally stable, may be realised by configurations with more than one macroscopically occupied site.  

Given $I(\rho,m)$, the canonical free energy and large deviations of the maximum under the canonical distributions are straightforward to compute
\begin{align}
  f(\rho)& =  \inf_{m\in [0,\rho]}I(\rho,m)\quad  \textrm{ and }  \nonumber\\
I_\rho(m) &:= -\lim_{L\to \infty}\frac{1}{L} \log  P_{L,N}\left( \cM_L = M\right)= I(\rho,m) - f(\rho)\ ,\label{eq:irhom}
\end{align}
where again $N/L\to\rho$ and $M/L\to m$. 
Note that $f(\rho )$ is simply a contraction over the most likely value for $\cM_L$ and gives the normalization of the rate function $I_\rho$. 

Below $\rho_c+a$ there is a unique minimum of $I_\rho(m)$ at $m=0$ which corresponds to the fluid phase.
Above $\rho_c+a$ there is another local minimum at $m = \rho - \rho_c$ which corresponds to the condensed state.
The fluid state exists for all densities $\rho$ and parameter values $a\geq 0,\ c\geq 1$, and is stable for $\rho <\rhotrans$ and metastable above (cf.\ Fig.~\ref{fig:ents}).
The transition density $\rho_{\rm trans}$ is then characterized by both local minima of $I_\rho (m)$ being of equal depth, i.e.
\begin{equation}
I_\rhotrans (0)=I_\rhotrans (\rho -\rho_c )=0\ .
\label{eq:rhotra}
\end{equation}
With the above results, this is equivalent to
\begin{equation}
f_{\rm fluid} (\rhotrans )=f_{\rm cond} (\rhotrans -\rho_c )+f_{\rm fluid} (\rho_c )\quad\mbox{with}\quad \rho_c =\frac{1}{c-1}\ .
\label{eq:rhotra2}
\end{equation}

\begin{figure}[tb]
\centering
{\mbox{\raisebox{24mm}{$(\mathrm{III}a)$}\includegraphics[height=0.2\textwidth]{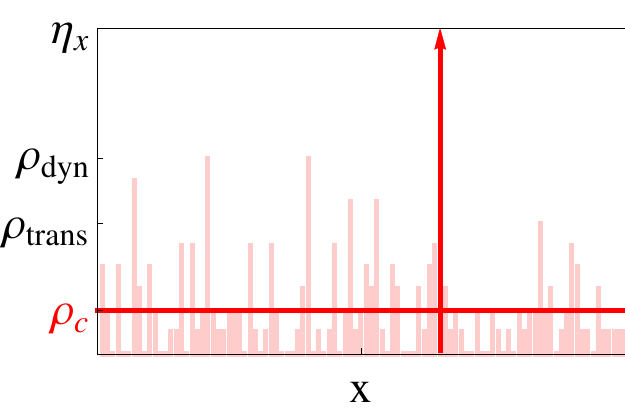}\raisebox{10mm}{$\ \to\!\!$}\includegraphics[height=0.19\textwidth]{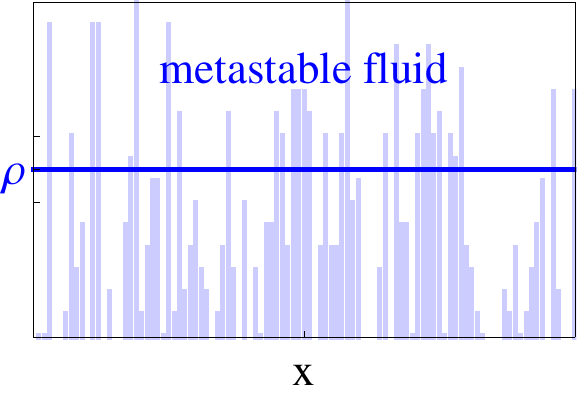}\raisebox{10mm}{$\ \to\ $}\includegraphics[height=0.19\textwidth]{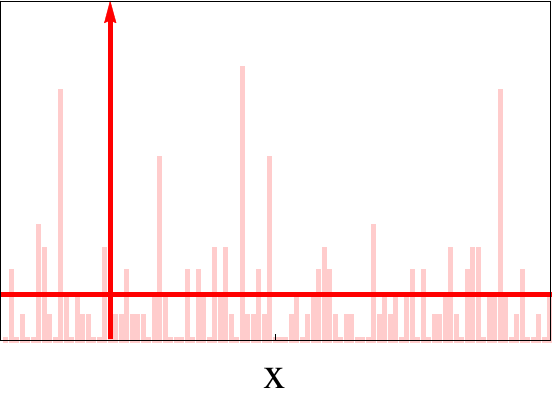}}}\\
{\mbox{\raisebox{24mm}{$(\mathrm{III}b)$}\includegraphics[height=0.2\textwidth]{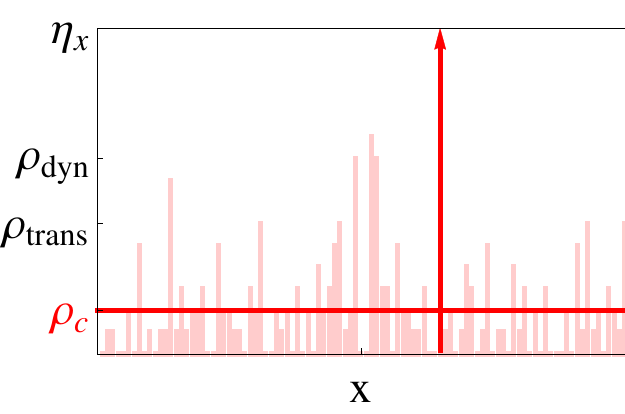}\raisebox{10mm}{$\ \to\ $}\includegraphics[height=0.19\textwidth]{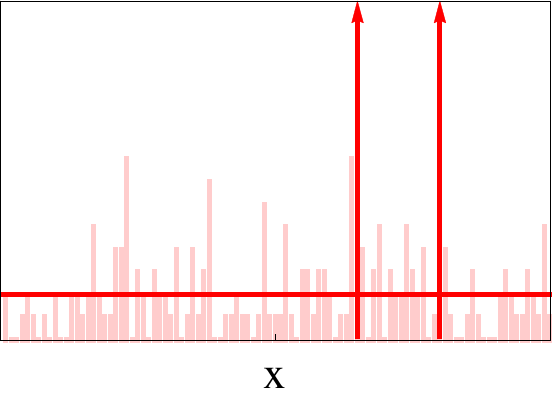}\raisebox{10mm}{$\ \to\ $}\includegraphics[height=0.19\textwidth]{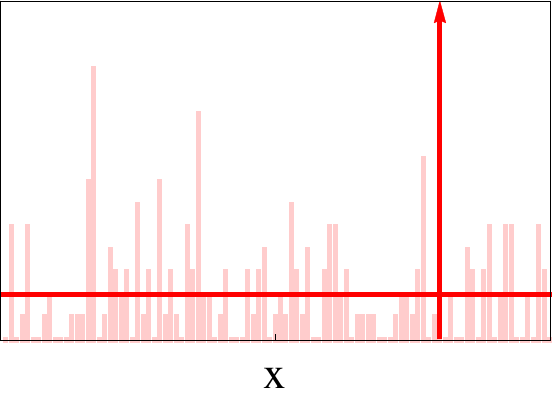}}}
\caption{\label{fig:CartoonMotion} Mechanisms of condensate motion. (IIIa) For $\rho<\rhodyn$, the condensate moves by losing particles to the bulk until the systems reaches a metastable fluid state. A condensate then reforms on a site chosen uniformly at random, independently of the previous location. (IIIb) For $\rho>\rhodyn$ 
the condensate moves by losing particles to another site and growing a second condensate. The system remains phase separated and the new condensate position depends on its initial location.} 
\end{figure}

\begin{figure}[t]
  \centering
  {\mbox{\raisebox{35mm}{$(\mathrm{III}a)$}\includegraphics[width=0.48\textwidth]{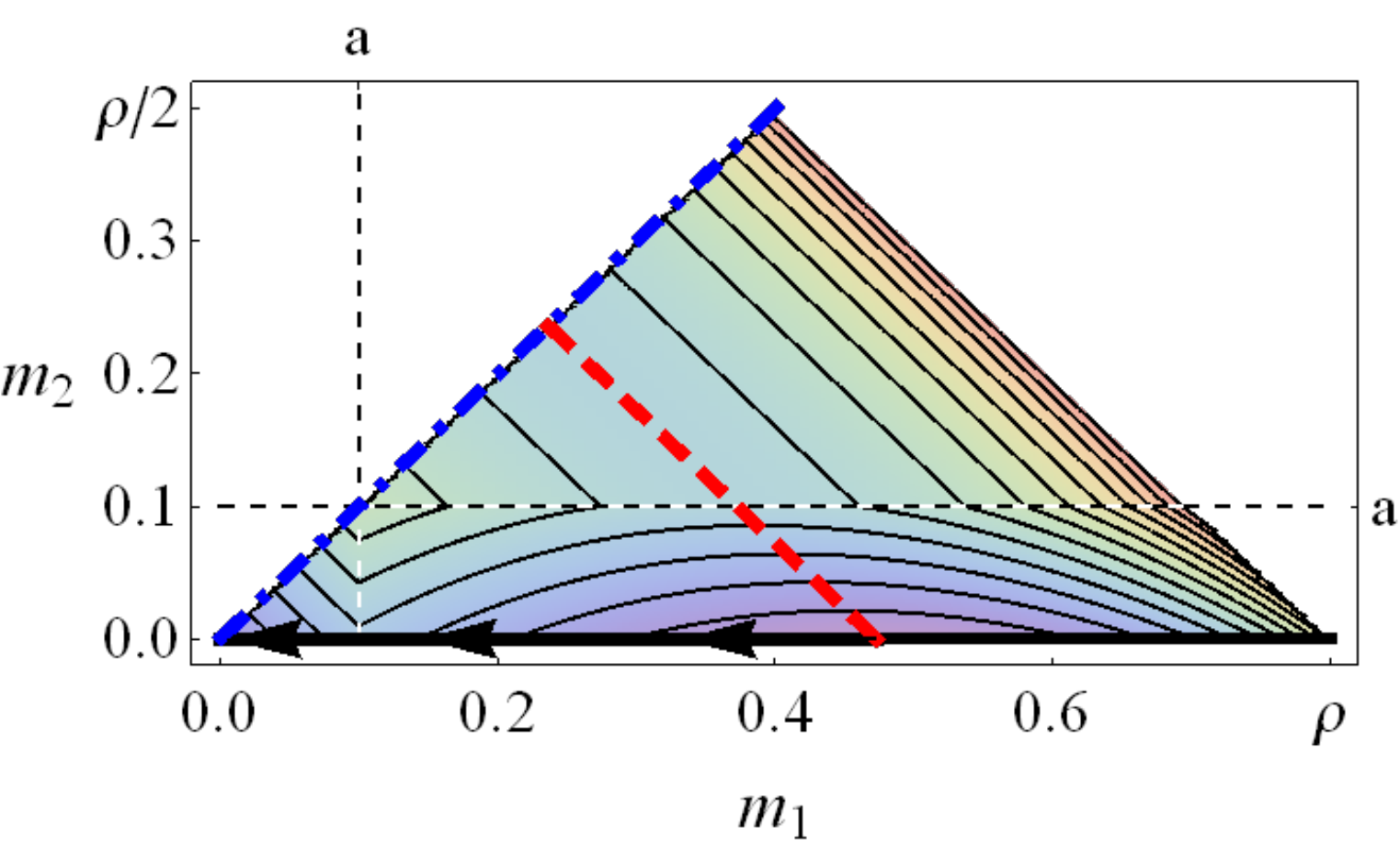}
  \includegraphics[width=0.42\textwidth]{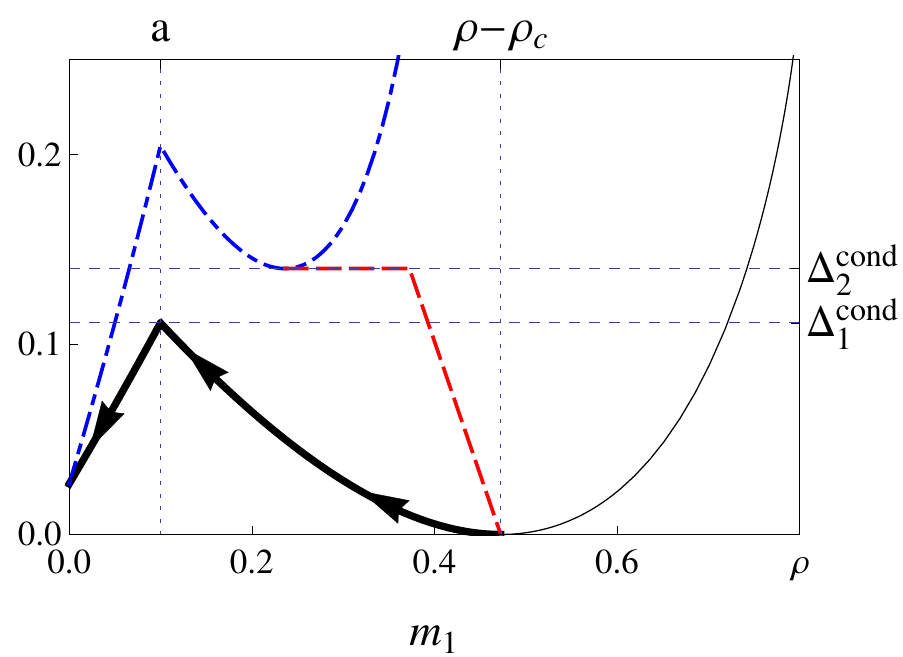}}}\\
  {\mbox{\raisebox{35mm}{$(\mathrm{III}b)$}\includegraphics[width=0.48\textwidth]{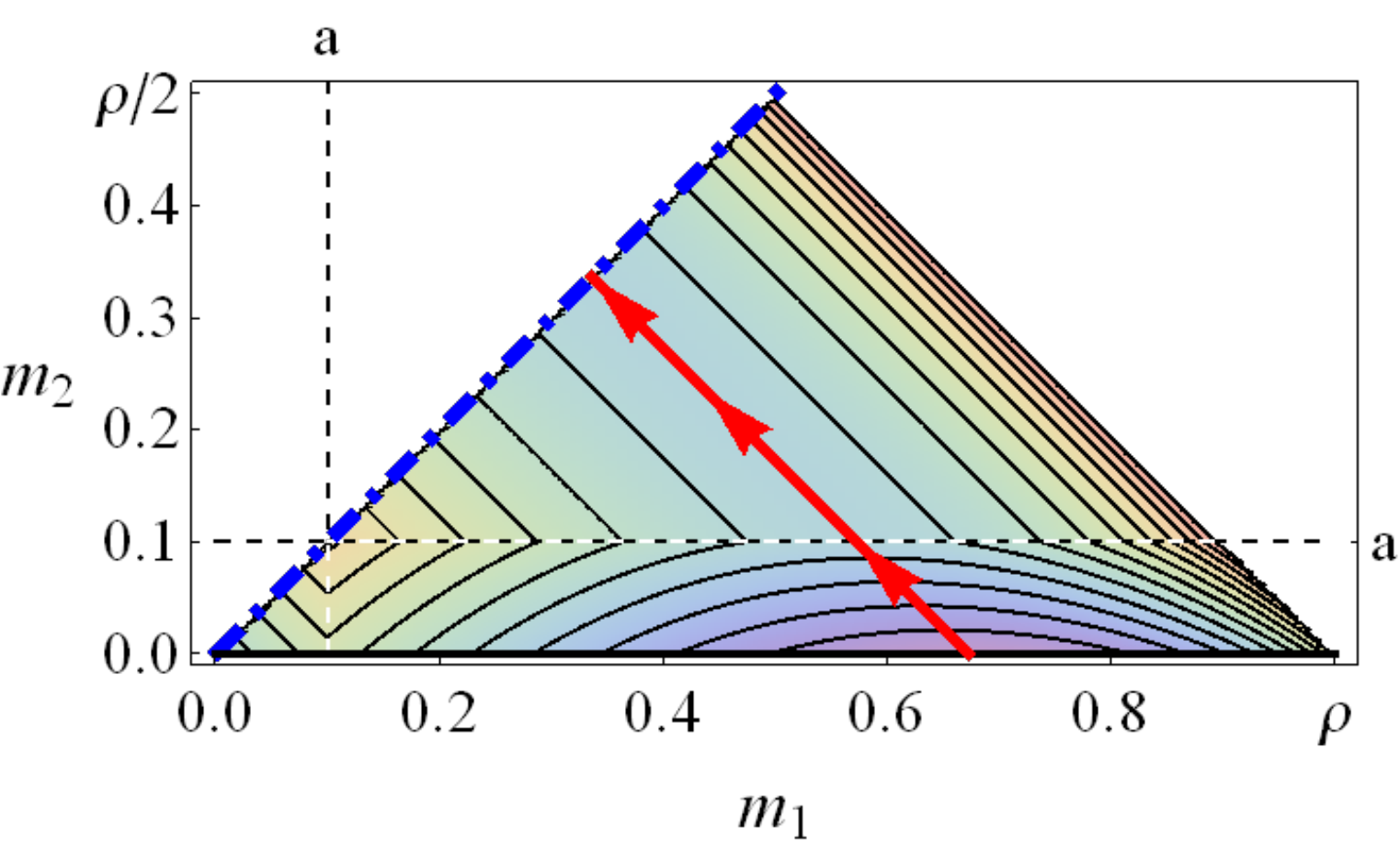}
  \includegraphics[width=0.42\textwidth]{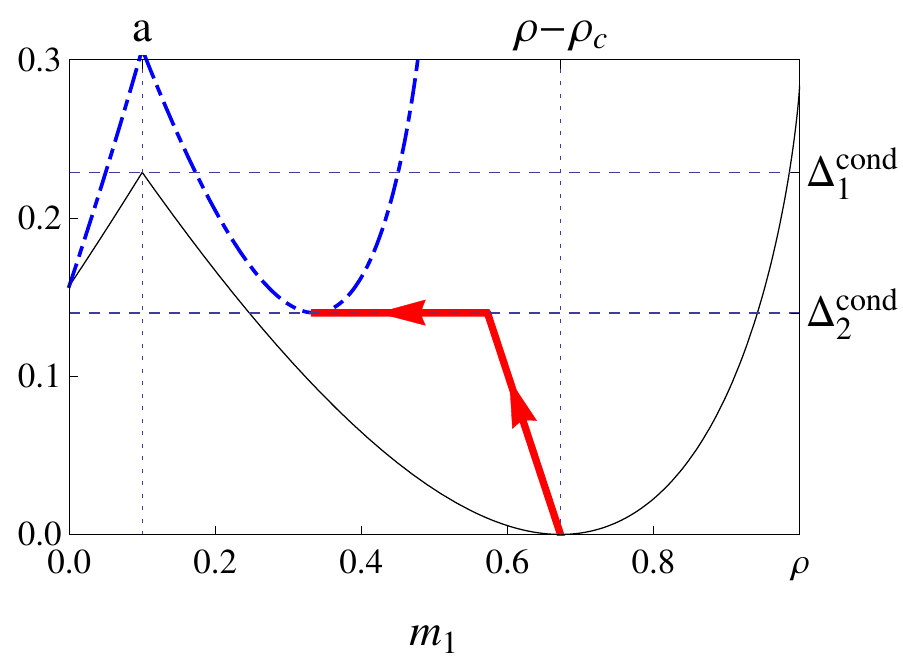}}}
	  \caption{\label{fig:MechA} Mechanisms for condensate motion from the free energy landscape, $a=0.1$, $c=4$. The top row shows mechanism (IIIa) with $\rho_{\rm trans}<\r=0.8 <\rho_{\rm dyn}$, the bottom row mechanism (IIIb) with $\rho=1 >\rho_{\rm dyn}$.
Left: Surface plot of $I^{(2)}_\rho(m_1,m_2)$ \eqref{eq:Itwo}, the arrowed line indicates the minimal action path to the diagonal which has to be reached for condensate motion. 
Right: The dashed blue line shows the landscape along the diagonal $I^{(2)}_\rho(x,x)$. The path along the x-axis $I^{(2)}_\rho (x,0)$ is shown as a full black line, and is chosen in mechanism (IIIa). The dashed red line shows the path to the diagonal by growing a second condensate with constant bulk density $\r_c$, chosen in mechanism (IIIb). $\Delta_1^{\rm cond},\Delta_2^{\rm cond}$ denote the respective exponential costs for the paths  \eqref{eq:Deltas}. }
\end{figure}

\noindent\textbf{The dynamic transition.}\\
Above $\rhotrans$ a typical stationary configuration is phase separated with the condensate on a single site.
Analogous to previous results \cite{beltranetal12}, due to translation invariance and ergodicity on large finite systems, the condensate will change location due to fluctuations. 
For large densities $\rho >\rhodyn$ the typical mechanism for this relocation is to stay phase separated and grow a second condensate (see Fig.~\ref{fig:CartoonMotion} IIIb), which is the same mechanism as identified in other super-critical ZRPs, see for example \cite{beltranetal12,AGL2}. 
This mechanism exhibits an interesting spatial depencence on the underlying random walk probabilities $p(x,y)$, which leads to a non-uniform motion of the condensate. 
For densities $\rhotrans <\rho <\rhodyn$ the typical mechanism is to dissolve the condensate and enter an intermediate metastable fluid state (see Fig.~\ref{fig:CartoonMotion} IIIa).
Since the system relaxes to a translation invariant metastable fluid state before the condensate reforms, the condensate reforms at a site uniformly at ramdom, independent of the geometry of the lattice.
This is very different from mechanism (IIIb) where the intermediate state is a saddle point with two condensates of equal height (cf. Fig. \ref{fig:MechA}). 
In both cases, the life time of intermediate states is negligible compared to the timescale on which the condensate moves and on this timescale the transition happens instantaneously.


To derive this transition, with the same approach as above we can calculate the canonical large deviations of the maximum and the second most occupied site $\cM_L^{(2)}$,
\begin{align}
\label{eq:Itwo}
I^{(2)}_\rho(m_{1},m_{2}) &= -\lim_{L\to\infty}\frac{1}{L}\log P_{L,N}[\cM_L = M_1,\ \cM_L^{(2)}=M_2] \nonumber\\
&= f_{\rm cond} (m_{1})  + I(\rho-m_{1},m_2) -f(\rho) \ ,
\end{align}
where $N/L\to\rho$, $M_1/L \to m_1$ and $M_2/L \to m_2$. 
This rate function essentially gives rise to a free energy landscape for the maximum and second most occupied site. 

In order for the condensate to move, the system must go via a state in which the maximum and second most occupied site differ in occupation by at most a single particle. In order to reach the diagonal $m_1 =m_2$ from a condensed state with $m_1 >0$ and $m_2 =0$ there are two relevant paths as shown in Fig. \ref{fig:MechA}. The first one is along the axis $m_2 =0$ towards the metastable fluid state with $m_1 =m_2 =0$ following the black line (mechanism IIIa), and the second one is along the red line with $m_1 +m_2 =\rho -\rho_c$, growing a second condensate reaching the diagonal at the  the local minimum of the blue curve, which is a saddle point in the full landscape (mechanism IIIb). The associated heights of the saddle points are given by
\begin{align}
\label{eq:Deltas}
 \Delta_1^{\textrm{cond}}(\rho) &:= I_\rho (a) - I_\rho (\rho- \rho_c)\nonumber\\
 \Delta_2^{\textrm{cond}}(\rho) &:= I^{(2)}_\rho (\rho-\rho_c -a,a) - I_\rho (\rho- \rho_c)=a \log c\ .
\end{align}
Plugging in (\ref{eq:fflu}) to (\ref{eq:irhom}) it is easy to see that $\Delta_1^{\textrm{cond}}(\rho)$ is increasing from $0$ for $\rho\geq \rho_c +a$ (see Fig.~\ref{fig:ents} right). Since $\Delta_2^{\textrm{cond}}(\rho)$ is constant, this implies that there is a dynamic transition at a density $\rhodyn$ characterized by
\begin{align}
\label{eq:rhodyn}
 \min\big\{\Delta_1^{\textrm{cond}}(\rho),\Delta_2^{\textrm{cond}}(\rho)\big\} = 
 \begin{cases}
 	\Delta_1^{\textrm{cond}}(\rho) & \textrm{for } \rho < \rhodyn\mbox{ (mech.~IIIa)} \\
 	\Delta_2^{\textrm{cond}}(\rho) & \textrm{for } \rho > \rhodyn\mbox{ (mech.~IIIb)} \ .
 \end{cases}
\end{align}
In this formalism we can also include
\begin{equation}
\Delta^{\textrm{fluid}}(\rho) := I_\rho (a) - I_\rho (0)
\label{eq:fdepth}
\end{equation}
as the depth of the fluid minimum, which provides another characterization of $\rhotrans$ as $\Delta^{\textrm{fluid}}(\rho)=\Delta_1^{\textrm{cond}}(\rho)$ as illustrated in Fig. \ref{fig:ents} on the right. After a straightforward computation this leads to
\begin{equation}
\rhodyn =\rhotrans +a\quad\mbox{and therefore}\quad \rhodyn >\rho_c +2a\ .
\label{eq:}
\end{equation}
Note that the saddle point at $m_1 = (\rho - \rho_c -a)$, $m_2 = a$ corresponding to $\Delta_2^{\textrm{cond}}$ only exists if $\rho > \rho_c + 2a$ and the system can sustain two macroscopically occupied sites. So while mechanism IIIa exists for all densities $\rho>\rho_c +a$ and therefore for $\rho >\rhotrans$, mechanism IIIb exists only for $\rho >\rho_c +2a$ (see Fig.~\ref{fig:ents} right). This is larger than $\rhotrans$ for $a$ large enough, but always below $\rhodyn$, when mechanism IIIb becomes typical.

\begin{figure}[t]
  \centering
  \mbox{
  \includegraphics[width=0.488\textwidth]{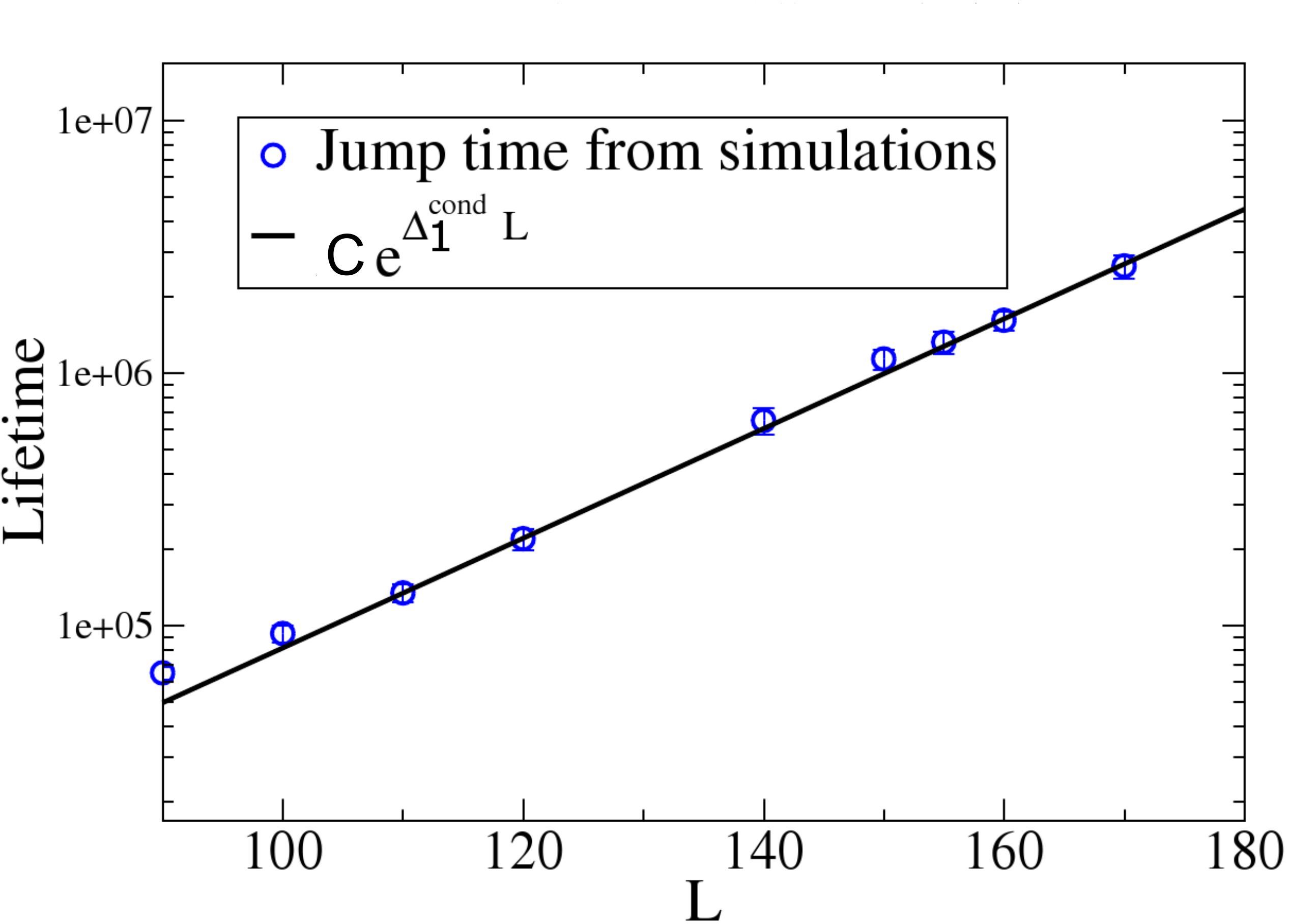}
  \hspace*{1ex}
  \includegraphics[width=0.488\textwidth]{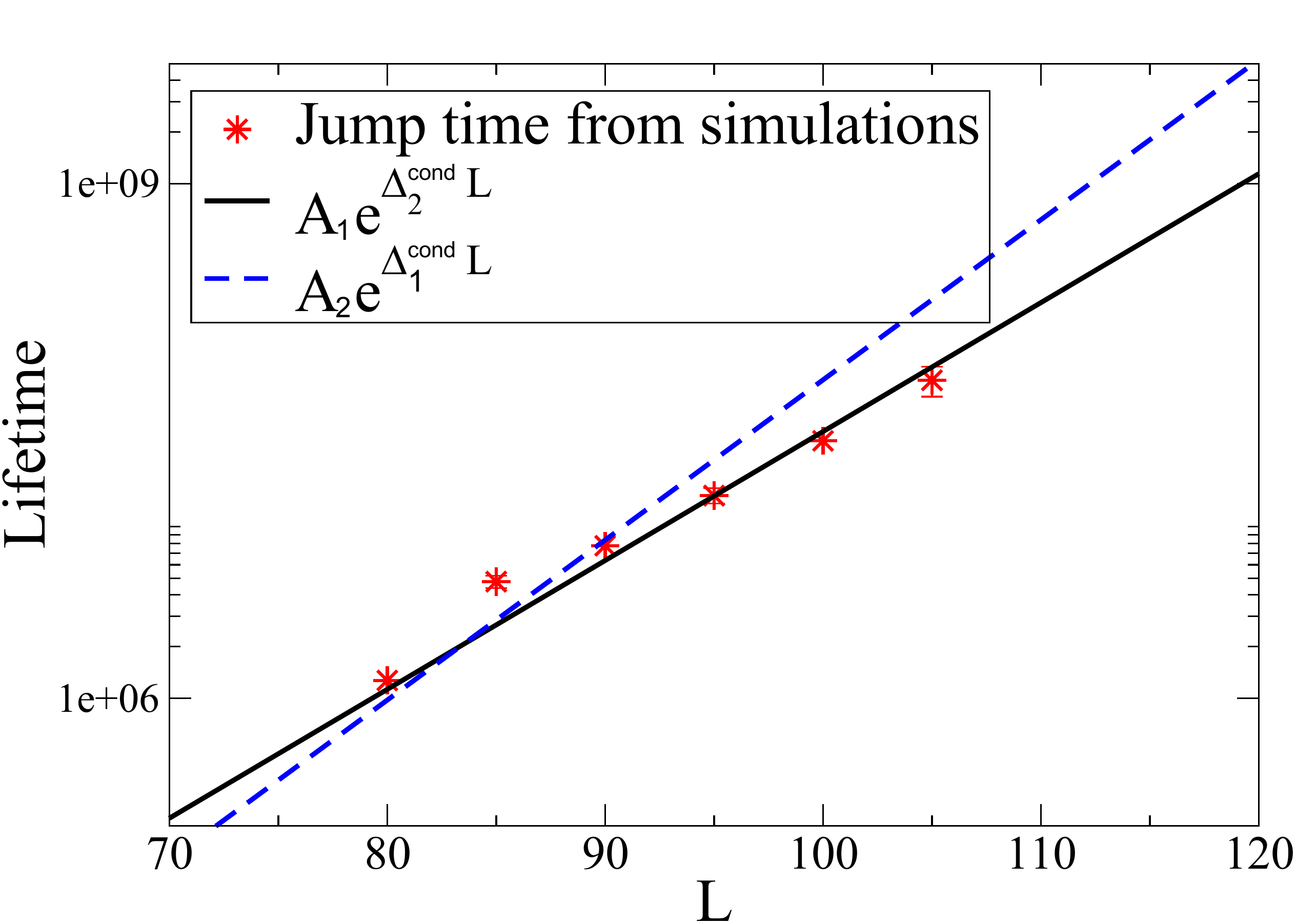}
  }
  \caption{\label{fig:MoveTimes}
	Average time for condensate motion as measured in simulations, with $c = 2$, $a=0.25$. Symbols show the average relocation time, sample standard deviation is of the size of the symbols. Left:  $\rho=1.75<\rhodyn$, mechanism $(\mathrm{III}a)$, relocation time is observed to grow with exponential factor $\Delta_1^{\textrm{cond}}$ \eqref{eq:Deltas}. Right: $\rho=2.4>\rhodyn$, mechanism $(\mathrm{III}a)$, relocation time grows with exponential factor $\Delta_2^{\textrm{cond}}$ \eqref{eq:Deltas}.}
\end{figure}

Although a complete rigorous description of the metastable motion of the condensate in this system is still an open problem, the exponential timescales associated with the corresponding activation times are directly related to the saddle point heights (as predicted by the Arrhenius law).
Also, the dynamics are expected to concentrate in the thermodynamic limit on the least action path (see \cite{freidlin,hugo14}). 
The expected lifetime of the fluid state, condensed state, and time to observe condensate motion are defined by
\begin{align}
T_{\textrm{fluid}}(\rho,L) &= \bbE_\r^{\textrm{fl}}\big[\inf \{t > 0 \mid \cM_L > a \}\big] \nonumber\\
T_{\textrm{cond}}(\rho,L) &= \bbE_\r^{\textrm{cd}}\big[\inf \{t > 0 \mid \cM_L < a \}\big]\nonumber \\
T_{\textrm{move}}(\rho,L) &= \bbE_\r^{\textrm{cd}}\big[\inf \{t > 0 \mid \cM_L =  \cM^{(2)}_L  \}\big]\ .
\end{align}
Here the expectations $\bbE_\r^{fl},\bbE_\r^{cd}$ are with respect to the dynamics with system size $L$ and density $\rho$ started from a configuration in the fluid and condensed states, respectively. 
The exponential growth of the life times with the system size is then related to the saddle point structure as follows,
\begin{align}
\label{eq:Ts}
\lim_{L\to \infty} \frac{1}{L} \log T_{\textrm{fluid}} &= \Delta^{\textrm{fluid}}(\rho),\quad 
\lim_{L\to \infty} \frac{1}{L} \log T_{\textrm{cond}} = \Delta_1^{\textrm{cond}}(\rho), \nonumber \\
\lim_{L\to \infty} \frac{1}{L} \log T_{\textrm{move}} &= \min\big\{\Delta_1^{\textrm{cond}}(\rho),\Delta_2^{\textrm{cond}}(\rho)\big\}\ .
\end{align} 
This behaviour and the dynamic transition are confirmed in simulations shown in Fig. \ref{fig:MoveTimes} for symmetric nearest neighbour dynamics in one dimension. 
In general, using the techniques of \cite{beltran13} the limiting motion of the condensate can be proved rigorously for reversible dynamics. 
For non-reversible ergodic dynamics the results are still expected to hold but are harder to prove, and additional restrictions may apply. 
First results on non-reversible condensate motion have just recently been achieved in \cite{landimasym,beltranetal12b}.


\begin{figure}[t]
  \centering
  
\end{figure}

\section*{Acknowledgments}
This work was partially supported by the Engineering and Physical Sciences Research Council
(EPSRC), Grant No.\ EP/E501311/1. P.C. would like to acknowledge the support of the University of Warwick IAS through a Global Research Fellowship.

%
%

\appendix
\section*{Appendix}

A standard approach for explicit computations of free energies is the use of grand-canonical distributions,
\begin{align*}
  \cP_{L,\mu}(\feta) = 
\frac{e^{\mu \sum_{x=1}^L\eta_x}}{\cZ_{L,\mu}}\overline P_L(\feta), \ \textrm{ with } \ \cZ_{L,\mu} =\langle e^{\mu \sum_{x=1}^L\eta_x} \rangle_{\overline P_L} = \left(\langle e^{\mu\cdot \eta_1} \rangle_{\overline P_L}\right)^L.
\end{align*}
The mean particle density is fixed by the conjugate parameter $\mu\in\bbR$ called the chemical potential. 
Note that the equality on the right hand side follows since the reference measure factorizes over lattice sites and the marginals on each site are identical.
The grand-canonical distributions are well defined for all $\mu \in (-\infty,1)$.
For fixed $L$, as $\mu \to 1$ the normalisation $\cZ_L(\mu)$ and the average particle density $\langle \eta_1 \rangle_{\cP_{L,\mu}}$ diverge.

The grand-canonical pressure is given by the point wise limit of the scaled cumulant generating function,
\begin{align}
 p(\mu) = \lim_{L\to \infty}\frac{1}{L}\log \cZ_{L,\mu} = 
  \begin{cases}
    \log\frac{c - e^{-1}}{c - e^{\mu-1}} & \textrm{ if } \mu < 1, \\
    \infty & \textrm{ if } \mu \geq 1.
  \end{cases}
\end{align}
The density can be computed as $R(\mu )=\partial_\mu p(\mu )$ and, as discussed in previous work \cite{stefangunter,CGreview}, the critical density is given by
\begin{equation}
\rho_c :=\lim_{\mu\nearrow 1} R(\mu )=\lim_{\mu\nearrow 1} \frac{e^{\mu -1}}{c-e^{\mu -1}} =\frac{1}{c-1} <\infty\ .
\label{eq:rhoc}
\end{equation}
Although $p(\mu)$ does not exist above $\mu_c = 1$ it can be extended analytically up to $1+ \log c$.
It turns out that this extended pressure is exactly the one associated to the grand-canonical distributions conditioned on no site containing more than $aL$ particles. 
These restricted grand-canonical distributions can be interpreted as metastable fluid states (see \cite{paulinprep} for details), and their pressure is given by  
\begin{align}
  p_{\rm fluid}(\mu) = \frac{e^\mu (c - e^{-1})}{c - e^{\mu-1}}\quad\mbox{where}\quad \mu <1 + \log c\ .
\end{align}
The free energy of the fluid  phase is then given by the Legendre-Fenchel transform of the pressure,
\begin{equation}
f_{\rm fluid} (\rho ):=\sup_{\m\in\bbR}[ \mu\rho - p_{\rm fluid} (\mu)]\ ,
\label{eq:fldef}
\end{equation}
which is explicitly given in (\ref{eq:fflu}). 
There are further interesting questions related to the equivalence of canonical and grand-canonical ensembles, which are discussed rigorously in \cite{paulinprep}.

\bibliographystyle{unsrt}
\bibliography{zrpmeta}
\end{document}